\documentclass[11pt]{article}

\usepackage{amsmath}
\usepackage{amssymb}
\usepackage{graphicx}
\usepackage{cite}
\usepackage{ulem} 
\usepackage{slashed}

\input{epsf}
\newcommand{\beq}{\begin{equation}}
\newcommand{\eeq}{\end{equation}}
\newcommand{\beqa}{\begin{eqnarray}}
\newcommand{\eeqa}{\end{eqnarray}}

\newcommand{\bea}{\begin{eqnarray}}
\newcommand{\eea}{\end{eqnarray}}

\begin{document}

\vspace{.9in}

\begin{center}

\bigskip

{{\Large\bf  New insights into the spin structure of the
    nucleon
}}

\end{center}

\vspace{.3in}

\begin{center}
{\large 
V\'eronique Bernard$^\ast$\footnote{email: bernard@ipno.in2p3.fr},
Evgeny Epelbaum$^\star$\footnote{email: evgeny.epelbaum@rub.de},
Hermann Krebs$^\star$\footnote{email: hermann.krebs@rub.de},
Ulf-G. Mei{\ss}ner$^\ddagger$$^\dagger$\footnote{email:
  meissner@hiskp.uni-bonn.de}
}

\vspace{1cm}

$^\ast${\it Groupe de Physique Th\'eorique, Institut de Physique Nucl\'eaire\\CNRS/Universit\'e Paris-Sud
  11, 91406 Orsay Cedex, France}

\bigskip

$^\star${\it Institut f\"ur Theoretische Physik II, Ruhr-Universit\"at Bochum,\\
  D-44780 Bochum, Germany}

\bigskip

$^\ddagger${\it Universit\"at Bonn,
Helmholtz--Institut f\"ur Strahlen-- und Kernphysik (Theorie) and\\
Bethe Center for Theoretical Physics, Universit\"at Bonn,
D-53115 Bonn, Germany}

\bigskip

$^\dagger${\it Forschungszentrum J\"ulich, Institut f\"ur Kernphysik, 
Institute for Advanced Simulation\\  and J\"ulich Center for Hadron Physics,
D-52425 J\"ulich, Germany}

\bigskip

\bigskip

\end{center}

\vspace{.4in}

\thispagestyle{empty} 

\begin{abstract}\noindent 
We analyze the low-energy spin structure
of the nucleon  in a covariant effective field theory 
with explicit spin-3/2 degrees of freedom to third order in the
small scale expansion. Using the available data on 
the strong and electromagnetic width of
the $\Delta$-resonance, we  give parameter-free
predictions for various spin-polarizabilities and
moments of spin structure functions. 
We find an improved description of the nucleon spin structure
at finite photon virtualities for some observables and
point out the necessity of a fourth order calculation.
\end{abstract}

\medskip


\smallskip
 

\vfill

\pagebreak

\section{Introduction}

 The internal spin structure of the nucleon which is 
described by structure functions in deep inelastic lepton-hadron scattering 
has been extensively studied in the last decades
both theoretically and experimentally, see e.g. \cite{Kuhn:2008sy} for a
review. The structure functions are related to the
real, virtual  or double virtual Compton scattering (${\rm V}^2{\rm CS}$) 
amplitudes by various sum rules which connect information at all energy scales.
On the experimental side only recently it has become possible to work with
polarized beams and polarized targets which is necessary for studying the
nucleon spin structure. One of the main goals of the Jefferson Lab activities
is to provide a precise experimental mapping of spin-dependent observables 
from low momentum transfer to the multi-GeV region, see e.g. 
\cite{Amarian:2002ar,Amarian:2004yf,Prok:2008ev} for the early  measurements
(that also cover the low-energy region). Concerning the
low-energy spin structure -- which is at the center of this investigation -- 
more data 
also at smaller photon virtualities $Q^2$ have been taken at Jefferson
Lab and their analysis will be completed soon, see e.g. Refs.~\cite{Chen:2010qc,CD12}. 
Therefore, it is timely to
reconsider the theoretical predictions for the moments of the nucleon spin
structure functions.

At very low energies, far below the chiral symmetry scale of the order of 
$1~{\rm GeV}$, the  nucleon dynamics is dominated by
chiral symmetry of QCD and for this reason can be rigorously
described by chiral perturbation theory (CHPT). This theory provides a
systematic expansion in low momenta and masses of the Goldstone bosons 
(identified with the pions in the case of two flavors). At low photon
virtualities one can make rigorous predictions for the spin-dependent part of 
${\rm V}^2{\rm CS}$ and use the Jefferson Lab data to test the 
chiral dynamics of QCD. 

${\rm V}^2{\rm CS}$ has already been considered within the CHPT framework up to
${\cal O}(q^4)$ in the chiral expansion by several theoretical groups, 
see~\cite{Bernard:2007zu} for a review (here, $q$ denotes a genuine small
parameter like external momenta or the pion mass). In standard CHPT, 
all effects of 
the 
$\Delta(1232)$-resonance degrees of freedom are encoded in
the low-energy constants. Chiral symmetry prevents spin-dependent 
counterterms of  ${\cal O}(q^3)$ and ${\cal O}(q^4)$ in ${\rm V}^2{\rm CS}$
such that their first contribution 
is possible at  ${\cal O}(q^5)$ in this scheme.
However, once one introduces the spin-3/2 degrees of freedom explicitly, 
one realizes that their tree-level contributions are large and far from 
being suppressed. This suggests to perform a systematic analysis of ${\rm
  V}^2{\rm CS}$ with explicit
deltas by using a covariant version of the so-called small scale
expansion (SSE) \cite{Hemmert:1997ye}. 
In that extension of CHPT, the nucleon-delta mass splitting, $\Delta =
m_\Delta -m_N$, is counted as an additional small parameter, thus the generic 
small parameter $\varepsilon$ collects external momenta, the pion mass and $\Delta$. 
Note that calculations within
the SSE
employing the heavy-baryon (HB) expansion 
were already performed by Kao et al.~\cite{Kao:2002cp}. 
Here, we address this issue up to the order $\varepsilon^3$ in a covariant
SSE. In contrast to the the covariant CHPT calculation  of 
Bernard et al.~\cite{Bernard:2002bs,Bernard:2002pw}, we do not use the method
of infrared regularization here as it leads to deformations of the analytical
structure at higher virtualities that leave a trace in the
$Q^2$-dependence
of certain observables (see also Ref.~\cite{Edelmann:1998bz}
for an early study of $\gamma_0(Q^2)$ using a relativistic version of baryon CHPT).
We obtain  parameter-free
predictions for various moments of the spin structure functions at low virtualities.
We focus, in particular, on the so-called forward and longitudinal-transverse
spin-polarizabilities, as these have posed particular problems to the
CHPT calculations. More precisely, the proton spin polarizability at the
photon point comes out larger in magnitude than experiment for most 
calculations  and also the magnitude of the $Q^2$-dependent neutron
longitudinal-transverse spin polarizability is predicted much smaller
than found in experiment. This latter finding is particularly puzzling
as the leading $\Delta$-resonance contributions are expected to cancel here. In this
paper, we will shed new light on these issues. It is also important to stress
that the contributions considered here are nothing but the leading order terms
in the chiral expansion of these spin polarizabilities based on an effective
Lagrangian of pions, nucleons, deltas and photons.

Our manuscript is organized as follows: In Sec.~\ref{sec:form1}, we
give the necessary formalism for double virtual Compton scattering
off the nucleon and the definition of the pertinent moments of spin
structure functions that are amenable to a low-energy expansion.
Sec.~\ref{sec:form2} is devoted to a short discussion of the
underlying effective Lagrangian, the covariant treatment of the
baryon fields and the pertinent Feynman diagrams to be calculated. 
Our results are discussed and presented in Sec.~\ref{sec:res}.
We end with a short summary and outlook in Sec.~\ref{sec:summ}.

\section{Formalism I: Double virtual Compton scattering}
\label{sec:form1}

The forward  tensor for double virtual Compton scattering 
in terms of the electromagnetic current $J^\mu$ is given by
\bea
T^{[\mu\nu]}&=&i\int d^4x\, e^{i q\cdot x}\langle P S|T J^\mu(x)J^\nu(0)|P
S\rangle~, \nonumber\\ J^\mu(x)&=&\sum_i e_i\bar{\psi}_i(x)\gamma^\mu\psi_i(x)~.
\eea
Here, $\psi_i(x)$ denotes a quark field of flavor $i$ with charge $e_i$, while 
$P$ and $S$ are the momentum and spin polarization of the nucleon,
respectively.  The spin-dependent
${\rm V}^2{\rm CS}$ tensor can be parameterized by two structure functions
\beq
T^{[\mu\nu]} =
-\frac{i}{2}\epsilon^{\mu\nu\alpha\beta}q_\alpha \, \biggl[S_\beta
S_1(\nu,Q^2) + \frac{1}{m_N^2}(P\cdot q S_\beta- S\cdot q P_\beta)S_2(\nu,Q^2)\biggr]~,
\eeq
which depend on two independent scalar variables $Q^2=-q^2$ (the photon
virtuality) and 
$\nu = P\cdot~q/m_N$ (the photon energy),  where $m_N$ is the nucleon mass. 
On the other hand, the differential cross
section of polarized spin-dependent inclusive lepton-nucleon scattering
(in the one-photon-exchange approximation) is proportional to the 
antisymmetric tensor
\beq
W^{[\mu\nu]} =\frac{1}{4\pi}\int d^4x\, e^{i q\cdot x}\langle P S|[J^\mu(x),J^\nu(0)]|P
S\rangle
\eeq
which can be parameterized again by two structure functions
\beq
W^{[\mu\nu]} = -\frac{i}{2}\epsilon^{\mu\nu\alpha\beta}q_\alpha\, \biggl[S_\beta
G_1(\nu,Q^2)
 +  \frac{1}{m_N^2}(P\cdot q S_\beta- S\cdot q P_\beta)G_2(\nu,Q^2)\biggr]~.
\eeq
The amplitudes for ${\rm V}^2{\rm CS}$  can be analytically continued to the 
complex $\nu-$plane. They have poles at $\nu_{c}=\pm Q^2/2m_N$, corresponding
to $s-$ and $u-$channel elastic scattering and two cuts on the real
axis extending from $\nu=\pm \nu_{c}$ to $\pm \infty$.
Using analyticity  and assuming a sufficient fast fall-off of
the structure functions at large $\nu$,
one can relate the corresponding structure
functions to each other by dispersion integrals~\cite{Ji:1999mr}
\bea
S_1(\nu,Q^2)&=&4\int_{Q^2/2m_N}^\infty \frac{dz \,z\, G_1(z,Q^2)}{z^2-\nu^2}~,\\
S_2(\nu,Q^2)&=&4\int_{Q^2/2m_N}^\infty \frac{dz \,\nu\, G_2(z,Q^2)}{z^2-\nu^2}~.
\eea
In the derivation of these relations crossing symmetry
\begin{equation}
S_1(-\nu,Q^2)=S_1(\nu,Q^2),\quad S_2(-\nu,Q^2)=-S_2(\nu,Q^2)~,
\end{equation}
has been  used. Usually one works with the inelastic version of the dispersion
relations where by inelastic we mean that the elastic part of the amplitudes
-- which have a pole at $Q^2/2m_N$ -- is subtracted. In this case the cut starts at
the pion production threshold $s=s_0=(m_N+M_\pi)^2$, which is equivalent to
$\nu_0=(Q^2+M_\pi^2)/(2m_N)+M_\pi$
The dispersion relations are then given by
\bea
\bar{S}_1(\nu,Q^2) &=& 4\int_{\nu_0}^\infty \frac{dz \,z\, G_1(z,Q^2)}{z^2-\nu^2}~,\\
\bar{S}_2(\nu,Q^2) &=& 4\int_{\nu_0}^\infty \frac{dz \,\nu\, G_2(z,Q^2)}{z^2-\nu^2}~,
\eea
with $\bar{S}_i(\nu,Q^2) = S_i(\nu,Q^2) -S_i^{\rm elastic} (\nu,Q^2)$.

For small photon energies, the ${\rm V}^2{\rm CS}$ amplitudes can be
expanded in powers of $\nu^2$:
\bea
\bar{S}_1(\nu,Q^2) &=& \sum_{i=0}^\infty \bar{S}_1^{(2i)}(0,Q^2)\nu^{2i}~, \\
\bar{S}_2(\nu,Q^2) &=& \sum_{i=0}^\infty \bar{S}_2^{(2i+1)}(0,Q^2)\nu^{2i+1}~.
\eea
The next-to-leading order coefficients (called the forward  and the
longitudinal-transverse spin-polariza\-bility, respectively) in this series 
are directly related to the moments of structure functions and can be measured. 
In this work, we concentrate on the low--energy region. The aforementioned
forward and longitudinal-transverse  polarizabilities are given by
\begin{eqnarray}
\gamma_0(Q^2)&=&\frac{1}{8\pi}\left(\bar{S}_1^{(2)}(0,Q^2)
   -\frac{Q^2}{m_N}\bar{S}_2^{(3)}(0,Q^2)\right),\\
\delta_0(Q^2)&=&\frac{1}{8\pi}\left(\bar{S}_1^{(2)}(0,Q^2) 
   +\frac{1}{m_N}\bar{S}_2^{(1)}(0,Q^2)\right).
\end{eqnarray}
They obviously can be described as dispersion integrals and can be rigorously 
calculated by CHPT at low virtualities. Using a dispersion representation one
has access to experimental data such that polarizabilities provide a testing ground
for chiral dynamics of QCD. Similar formulae can be given for the generalized
GDH integral $I_A (Q^2)$ and the first moments of the spin structure 
functions $\Gamma_1(Q^2)$, see Ref.~\cite{Ji:1999mr}. For completeness, 
we give the corresponding expressions:
\beqa
I_A(Q^2) = \frac{m_N^2}{4e^2}\left[ \bar{S}^{(0)}_1(0,Q^2) - 
\frac{Q^2}{m_N} \bar{S}^{(1)}_2(0,Q^2) \right]~, \nonumber\\
\Gamma_1(Q^2) = \frac{Q^2}{2m_N^2} I_1(Q^2)~, ~~ I_1(Q^2) = \frac{m_N^2}{4e^2}
 \bar{S}^{(0)}_1(0,Q^2)~.
\eeqa
These observables will also be considered here.

\section{Formalism II: Effective Lagrangian and one-loop calculation}
\label{sec:form2}

We now consider the underlying chiral Lagrangian. In
Ref.~\cite{Bernard:2002pw}, the chiral pion-nucleon Lagrangian in the
presence of external sources,  ${\cal L}_{\pi N} [U, N, \bar N; s, p, v_\mu,
a_\mu]$, was utilized combined with infrared regularization to separate the soft 
(long-range) from the hard (short distance) dynamics. Contributions from the 
$\Delta(1232)$ resonance at tree level  were added in a phenomenological 
approach and shown to be important. Here, we improve this calculation by 
extending the underlying effective field theory (EFT) to include the delta 
based on the so-called covariant small scale expansion to 
${\cal O}(\varepsilon^3)$. We use the explicit form of the spin-3/2 propagator
from Ref.~\cite{Bernard:2005fy}. Also, we do not use infrared regularization
as done in Ref.~\cite{Bernard:2002pw} but rather utilize dimensional
regularization. For the case at hand, this is a consistent scheme
as counter terms in the four-point function
only show up at  ${\cal O}(\varepsilon^5)$.
Therefore,
no power-counting violating contributions appear up-to-and-including 
${\cal O}(\varepsilon^4)$ and the corresponding loop corrections to
V$^2$CS are all finite after mass and coupling constant renormalization.
Note, however, that one has to deal with  some LECs in the three-point 
functions that appear as parts of the fourth order diagrams. In case 
of nucleon intermediate states, these are nothing but the 
anomalous magnetic moment of the proton and the neutron, see also 
Ref.~\cite{Bernard:2002pw}. In case of delta intermediate
states, we have in addition dimension two LECs from ${\cal L}^{(3)}_{\pi N
  \Delta}$, that can be fixed from $\Delta \to N$ transition form factors.

The chiral Lagrangian for V$^2$CS in the pion-nucleon sector is given in
Ref.~\cite{Bernard:2002pw}. The pertinent new Lagrangian structures related to 
the inclusion of the spin-3/2 fields read (for the construction principles, 
see \cite{Hemmert:1997ye})
\bea
{\cal L}_{\pi N\Delta}^{(1)} &=& h_A \, \bar\psi_i^\mu \, \omega_\mu^i \, N +
{\rm h.c.}~, \\
{\cal L}_{\pi\Delta\Delta}^{(1)} &=& \bar\psi_i^\mu
\,(i\slashed{D}_{\mu \nu}^{ij}  - m_\Delta \gamma_{\mu\nu} \delta^{ij}) \, \psi_j^\nu~,\\
{\cal L}_{\pi N \Delta}^{(2)} &=& \frac{1}{2}b_1 \, \bar\psi_i^\mu \,
if_{+\mu\alpha}^i\gamma^\alpha\gamma_5 \, N +
{\rm h.c.}~, 
\eea
with
\bea
\slashed{D}_{\mu\nu}^{ij} &=& \gamma_{\mu\nu\alpha}\,D_{ij}^\alpha~,~~
D_{ij}^\alpha = (\partial^\alpha + \Gamma^\alpha) \, \delta_{ij} 
- i\epsilon_{ijk}\langle\tau^k\Gamma^\alpha\rangle~,\nonumber\\
\Gamma^\alpha &=& \frac{1}{2}[u^\dagger, \partial_\mu u] - \frac{i}{2}
u^\dagger (v_\mu+a_\mu) u - \frac{i}{2} u (v_\mu - a_\mu)u^\dagger\nonumber\\
f_{+\mu\alpha}^i &=& \frac{1}{2}\langle\tau^i f_{+\mu\nu}\rangle~,~~
\omega_\mu^i = \frac{1}{2}\langle\tau^i u_\mu \rangle~,~~\nonumber\\
\gamma_{\mu\nu\alpha} &=& \frac{1}{4}\bigl\{[\gamma_\mu , \,  \gamma_\nu],
\gamma_\alpha\bigr\}~,~~~~~
\gamma_{\mu\nu} = \frac{1}{2} [\gamma_\mu , \,  \gamma_\nu]
~.
\eea
Here, $\psi_i^\mu$ is a conventional Rarita-Schwinger spinor for the spin-3/2
fields, $N$ denotes the nucleon bi-spinor (throughout, we work in the isospin
limit), $h_A$ is the leading $\pi N\Delta$ axial-coupling (analogous to $g_A$ in
the pion-nucleon sector) and $b_1$ is the leading photon-nucleon-delta
coupling of chiral dimension two (much like the nucleon magnetic moment that
appears first in ${\cal L}_{\pi N}^{(2)}$). As usual, the pions are collected
in the matrix-valued field $U(x) = u^2(x)$.
We only need the external vector source $v_\mu =  Q A_\mu$, with $A_\mu$ the
photon field and $Q =(1,0)e$ the nucleon charge matrix. Therefore 
$f_{+\mu\nu} = F_{\mu\nu}(u Q u^\dagger+ u^\dagger Q u)$.

\begin{figure}
\begin{center}
\includegraphics[width=0.65\textwidth]{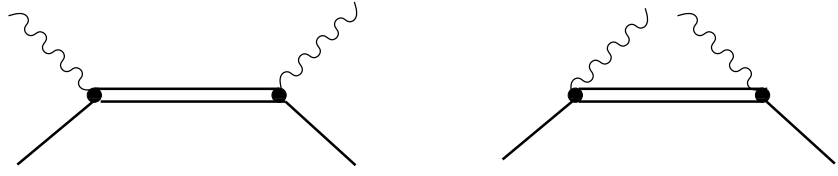}
\end{center}
\caption{Tree diagrams at ${\cal O}(\varepsilon^3)$. Solid, double
  and wiggly lines denote nucleons, deltas and photons, in
  order. The filled circle is an insertion from ${\cal L}_{\pi N \Delta}^{(2)}$.}
\label{fig:tree}
\end{figure}

\begin{figure}
\begin{center}
\includegraphics[width=0.85\textwidth]{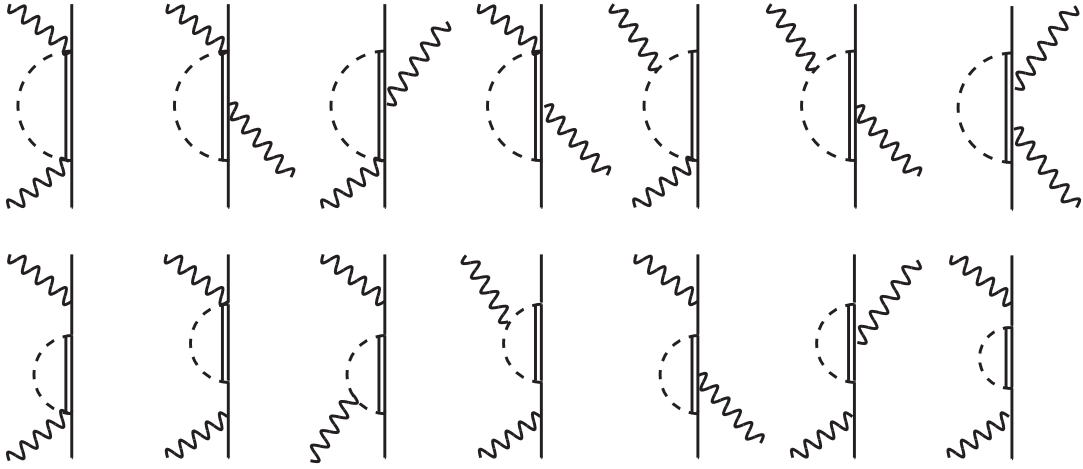}
\end{center}
\caption{Delta-loop diagrams at ${\cal O}(\varepsilon^3)$. Solid, double,
  dashed and wiggly lines denote nucleons, deltas, pions and photons, in
  order. Crossed graphs are not shown.}
\label{fig:diag}
\end{figure}

Based on this, we are now in the position to calculate V$^2$CS at low
energies in the covariant SSE. At order ${\cal O}(\varepsilon^3)$, 
we have tree and the leading one-loop
graphs involving the $\Delta$-resonance, see Figs.~\ref{fig:tree} and 
\ref{fig:diag},  respectively. The corresponding third order pion-nucleon
loop graphs are e.g. displayed in Fig.~1 of Ref.~\cite{Bernard:2002pw}.
Note that since the delta propagator is of  ${\cal O}(\varepsilon^{-1})$, the tree
graphs with two insertions from ${\cal L}_{\pi N \Delta}^{(2)}$ appear first
at third order in the SSE. Note further that at this order there are no unknown
low-energy constants (LECs), since the couplings $h_A$ and $b_1$ can be
determined from the decays $\Delta \to N\pi$ and $\Delta \to N\gamma$,
respectively. More precisely, the strong width of the $\Delta$ is given
in terms of the LEC $h_A$ as
\beq\label{eq:Dws}
\Gamma_\Delta^{\rm str} = h_A^2 \frac{((m_\Delta-m_N)^2-M_\pi^2)^{3/2}
((m_\Delta+m_N)^2-M_\pi^2)^{5/2}}{192\, F_\pi^2 \,\pi \,m_\Delta^5} = (118\pm
2)~{\rm MeV}~,
\eeq
and similarly for the electromagnetic width in terms of $b_1$
\beq\label{eq:Dwe}
\Gamma_\Delta^{\rm em} = e^2 b_1^2
\frac{(m_\Delta^2-m_N^2)^3(3m_\Delta^2+m^2)}{576\,\pi\, m_\Delta^5}~,
\eeq
with $\Gamma_\Delta^{\rm  em}/(\Gamma_\Delta^{\rm  em} +  \Gamma_\Delta^{\rm
  str}) = 0.55 \ldots 0.65\%$.
The predictions for the generalized  spin polarizabilities and other moments
of the spin structure functions are thus {\it parameter-free}. 
It is also important to stress
that in the covariant scheme employed here one has more loop diagrams at leading
order as compared to the heavy baryon approach (cf. Fig.~2 in
\cite{Kao:2002cp}). In that approach,  the ``missing''
graphs only appear at fourth order
due to the additional counting in the inverse baryon mass. Here, we have
to deal with 14 different topologies as shown in Fig.~\ref{fig:diag}. None
of  them
involves the leading,  dimension-two $\Delta N \gamma$-vertices, 
such contributions only start at ${\cal O}(\varepsilon^4)$. Still, the algebra 
to evaluate the diagrams shown in  Fig.~\ref{fig:diag} is
non-trivial. In particular the box-type diagrams which three spin-3/2
propagators generate a large number of terms, as the 
spin-3/2 field propagator is given by 
\beq
S^{\mu\nu} = \frac{p\!\!/ + m_\Delta}{p^2-m^2_\Delta}\left( -g^{\mu\nu}
+\frac{1}{3} \gamma^\mu\gamma^\nu + \frac{1}{3m_\Delta}\left(\gamma^\mu p^\nu
- \gamma^\nu p^\mu\right) +   \frac{2}{3m_\Delta^2} p^\mu p^\nu \right)~.
\eeq
For example, the box diagram (right-most graph in the upper row of 
Fig.~\ref{fig:diag}) has $5^3 = 125$ times more terms than the corresponding
pion-nucleon box graph.  Therefore, we have developed 
our own algebraic program that combines FORM \cite{Vermaseren:2000nd}
and Mathematica to calculate the tree and the loop diagrams. The code 
is able to reduce tensor integrals of any rank in the relativistic and
the heavy baryon formalism. FORM is used to reduce the pertinent tensor
integrals with higher powers of propagators and shifted dimensions while
Mathematica is utilized to perform the standard Passarino-Veltman reduction
\cite{Passarino:1978jh} (if required). In particular, the program 
allows for an easy heavy mass
reduction of any given relativistic formulation.

We have calculated the spin-polarizabilities $\gamma_0 (Q^2)$, 
$\delta_0 (Q^2)$, the generalized GDH integral $I_A(Q^2)$ and also the
first moment $\Gamma_1(Q^2)$ for the neutron and the proton. The resulting expressions
for the loop contributions are very lengthy and will not be given here 
explicitly\footnote{They can be
made available as a Mathematica notebook upon request from Hermann Krebs.}. 
However, we mention that our framework allows to take the heavy baryon 
limit in which the nucleon and the delta are considered as heavy, static 
sources keeping the mass splitting fixed. Indeed, we recover the heavy 
baryon results of~Ref.\cite{Kao:2002cp}. For better comparison, 
we give here  the explicit Born terms corresponding to Fig.~\ref{fig:tree}
\beqa\label{eq:born}
S_1^{\Delta-{\rm Born}}(\nu, Q^2)&=&\frac{2 e^2 b_1^2}{9
  m_\Delta^2}\cdot\frac{1}{(m_\Delta^2-m_N^2 + Q^2)^2-4 m_N^2 \nu^2}
\bigg[2 m_N^2 \nu^2
  (m_N^2-3(Q^2+m_\Delta^2)) \nonumber\\
&+&Q^2(m_\Delta^2-m_N^2+Q^2)(3 m_\Delta^2-2
  m_N^2 + 2 Q^2 - 2 m_N m_\Delta)\bigg],\nonumber\\
S_2^{\Delta-{\rm Born}}(\nu,Q^2)&=&-\frac{4 e^2 b_1^2 m_N^2 \nu}{9
m_\Delta^2}\,\frac{(m_N+m_\Delta)(m_N^2+Q^2-2m_N m_\Delta)}{(m_\Delta^2-m_N^2 + Q^2)^2-4
m_N^2 \nu^2}~.
\eeqa

\section{Results and discussion}
\label{sec:res}

For obtaining numerical results, we use the following 
set of parameters: $g_A=1.27$, $F_\pi=92.21\,$MeV, 
$M_\pi=138.04$~MeV, $m_N=938.9\,$MeV, $\kappa_v=3.706$,
$\kappa_s=-0.120$, $m_\Delta=1232\,$MeV (in the appendix,
we also discuss some results obtained using the S-matrix pole mass
as determined e.g. in pion-nucleon scattering). 
For the $\Delta$ couplings, we obtain from Eqs.~(\ref{eq:Dws},\ref{eq:Dwe})
\beq\label{eq:delcoups}
h_A = 1.43 \pm 0.02~, ~~~~ b_1 = -(4.98\pm 0.27)/m_N~.
\eeq
For comparison, the corresponding large-$N_C$ relations yield
$h_A  = (3g_A)/(2\sqrt{2}) = 1.35$, and
$b_1 = -3(1+\kappa_p-\kappa_n)/({2\sqrt{2}m_N}) 
= -{5.0}/{m_N}$, which are consistent with the empirical values.
Note that we take the sign of $h_A$ and $b_1$ to be consistent
with the large-$N_C$ relations, as the formulae for the corresponding 
width are quadratic in these couplings.
We will generate theoretical errors by varying these couplings
within the ranges given above. Uncertainties due to neglected higher
orders will not be considered.

First, we consider the forward and the longitudinal-transverse spin-polarizabilities 
at the photon point, $\gamma_0 (0)$ and $\delta_0(0)$, respectively. 
We find using the central values of the input parameters
\beqa\label{eq:parts}
\gamma_0^p&=&2.07_{q^3}\, - \, 3.65_{\epsilon^3, {\rm
    tree}}\, - \, 0.16_{\epsilon^3, {\rm loop}} \, = \, -1.74
~~[\pm 0.40]~,\nonumber \\
\gamma_0^n&=&3.06_{q^3}\, - \, 3.65_{\epsilon^3, {\rm
    tree}}\, - \, 0.18_{\epsilon^3, {\rm loop}} \, = \, -0.77
~~[\pm 0.40]~,\nonumber\\
\delta_0^p&=&1.54_{q^3}\, - \, 0.36_{\epsilon^3, {\rm
    tree}}\, + \, 1.22_{\epsilon^3, {\rm loop}} \, = \,\,\,\,\,\, 2.40
~~[\pm 0.01]~,\nonumber \\
\delta_0^n&=&2.41_{q^3}\, - \, 0.36_{\epsilon^3, {\rm
    tree}}\, + \, 0.33_{\epsilon^3, {\rm loop}} \, = \,\,\,\,\,\, 2.38
~~[\pm 0.03]~,
\eeqa
in units of  $10^{-4}\,$fm$^4$. The first  term refers to the
third order pion-nucleon loop result, whereas the second and third
term are the delta tree and loop corrections at third
order in the SSE. In brackets, we give the results due to the
variation of $h_A$ and $b_1$ within the bounds given above.
We do not attempt here to estimate the error stemming  from 
the fourth (and higher) order terms -- this issue will be dealt
with in the future when we present the results of the complete
one-loop analysis.
As already found in Ref.~\cite{Bernard:2002pw}, the corrections from
tree-level delta graphs are large in the forward spin-polarizabilities
whereas the delta loop corrections for $\gamma_0^{n,p}$ are very 
small.\footnote{In this work we use dimensional and {\it not} 
  infrared regularization as in
  Ref.~\cite{Bernard:2002pw}. For this reason only qualitative
  comparison  is possible  between our results and that of~\cite{Bernard:2002pw}.}
This is different for the transverse-longitudinal polarizabilities,
where the tree contributions are suppressed (as it was also 
found in the heavy baryon calculation of~\cite{Kao:2002cp}).
We note that the parameter-free prediction for $\gamma_0^p$ 
agrees within 1.5~$\sigma$ with the empirical number, 
$\gamma_0^p = -1.00 \pm 0.08 \pm 0.12$ \cite{Dutz:2003mm}.
We note that the latter number is obtained using the well-known sum rule
for $\gamma_0$ in terms of the measured difference of the 
photon-proton cross sections with helicity $1/2$ and $3/2$ 
for photon energies between 200 and 1800~MeV combined with the 
MAID2003 prediction for the region between the threshold and
200~MeV.  This is a clear improvement as compared to earlier 
calculations employing either HBCHPT with explicit deltas or the
covariant ${\cal O}(q^4)$ calculation adding tree-level $\Delta$-contributions.
Of course, before one can claim success, one must consistently
evaluate the fourth order contributions from nucleon and delta
intermediate states. We also note the marked difference in the
delta-loop contribution to $\delta_0^p$. While in the heavy baryon
scheme this contribution is small, it is sizeable in our relativistic
approach. This can be 
largely
traced back to the box diagram (the right-most
 diagram in the upper row of Fig.~\ref{fig:diag}). As $m_N$ and $m_\Delta$ (with the
splitting fixed) tend to infinity, the contribution from this
diagram vanishes, whereas its value is $1.32 \cdot 10^{-4}$~fm$^4$
in our case. We have analyzed the $1/m_N$-expansion\footnote{The
$1/m_N$ expansion with fixed $\Delta=m_\Delta-m_N$ leads to the usual 
heavy baryon SSE.} of the box diagram. We find that
it only starts to contribute at ${\cal O}(1/m_N^2)$ and that the convergence
of the series is very slow. The effect is much more dramatic for the 
proton than for the neutron due to a much larger prefactor. This shows that
the heavy baryon expansion does not provide a good approximation to the 
covariant result for this observable.

\begin{figure}[t!]
\begin{center}
\includegraphics[width=0.33\textwidth,angle=270]{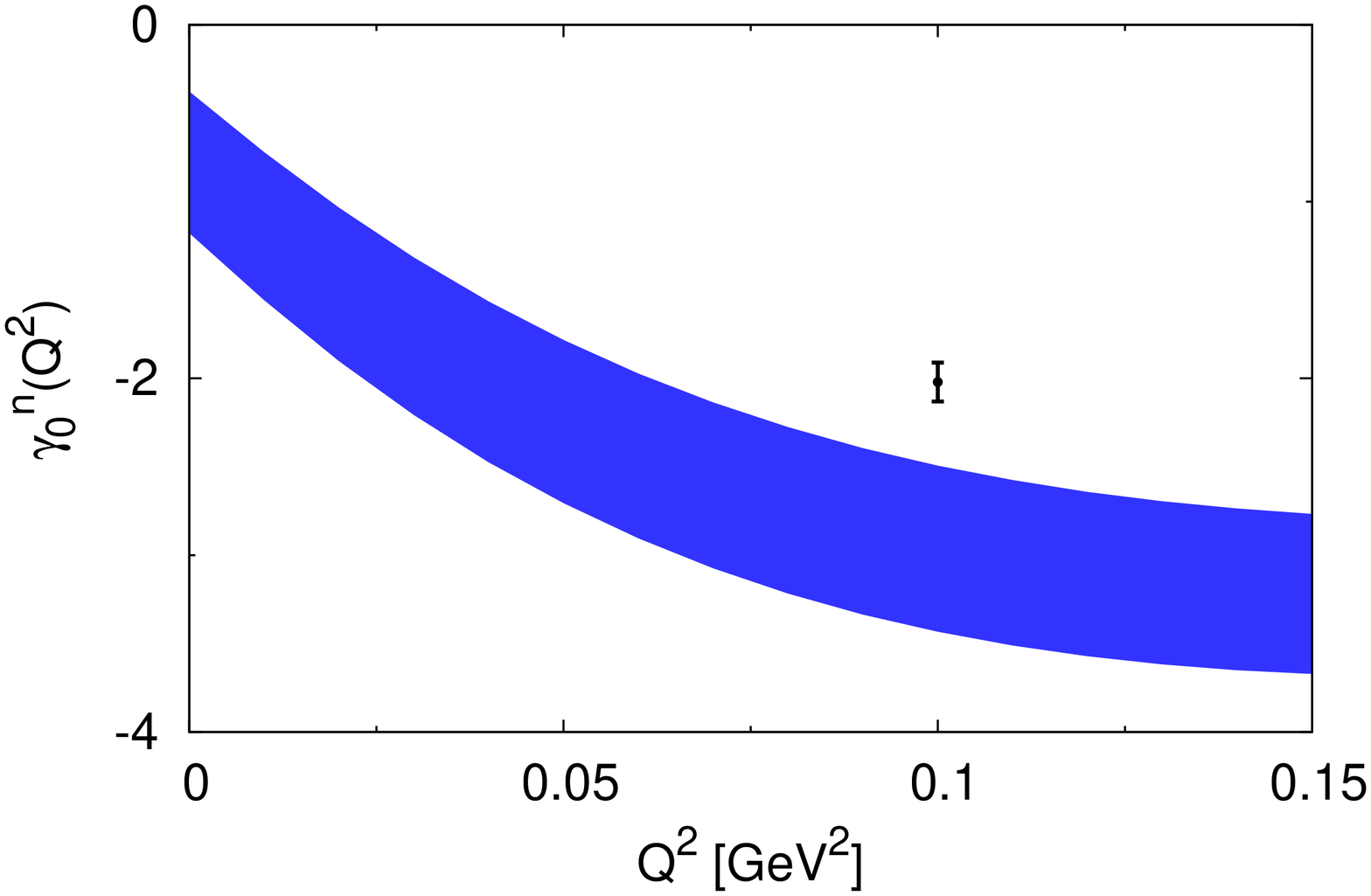}\hspace{0.5cm}
\includegraphics[width=0.33\textwidth,angle=270]{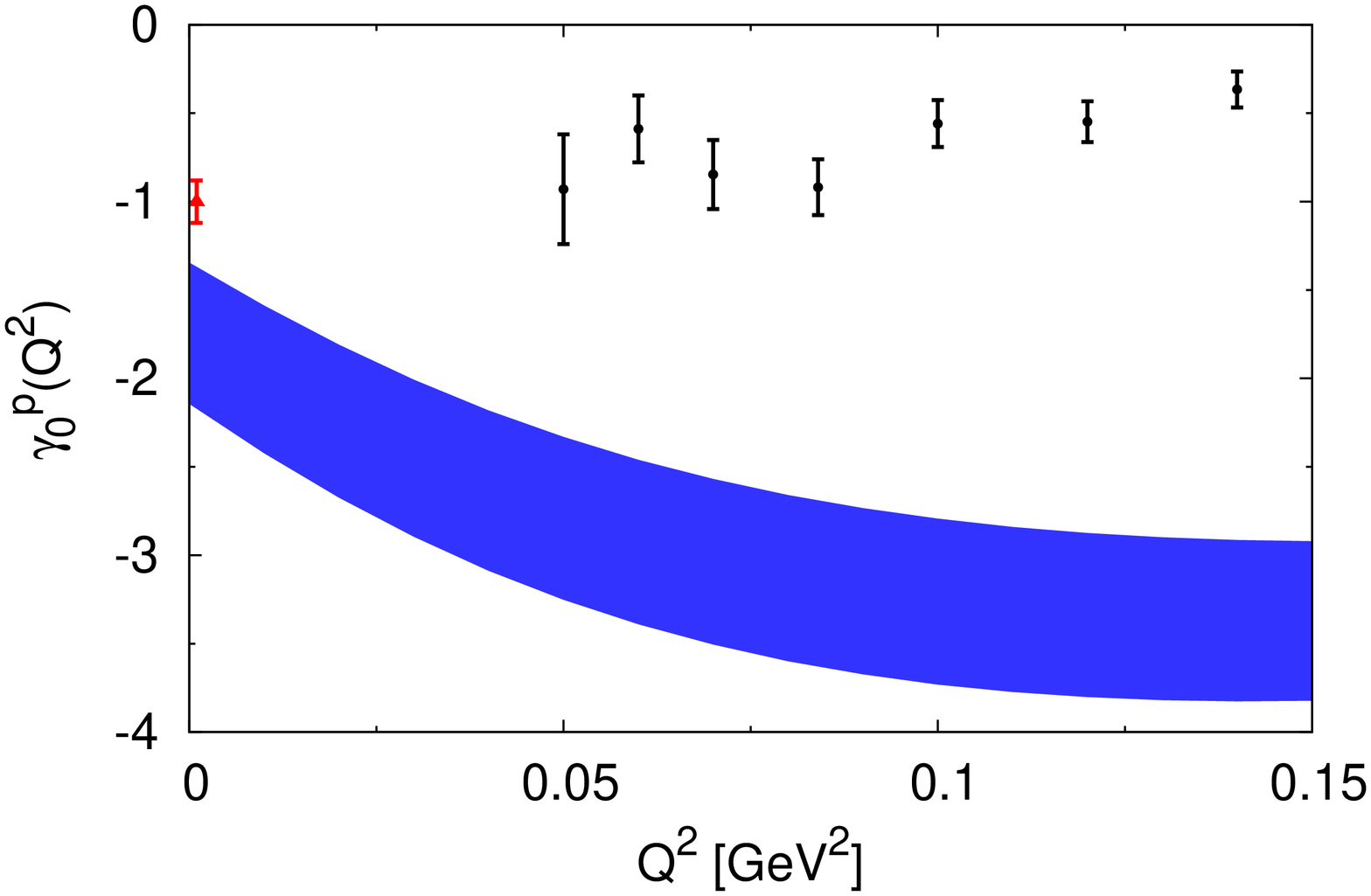}
\end{center}
\caption{Forward spin polarizability 
in units of $10^{-4}$ fm$^4$ at finite photon virtuality for the
neutron (left) and the proton (right). Neutron data:
Ref.~\cite{Amarian:2004yf} 
and proton data from Ref.~\cite{Dutz:2003mm} ($Q^2 =0$) and 
Ref.~\cite{Prok:2008ev}  ($Q^2 > 0$). Only statistical errors are shown.}
\label{fig:gamma0Q2}
\end{figure}

\begin{figure}[h!]
\begin{center}
\includegraphics[width=0.33\textwidth,angle=270]{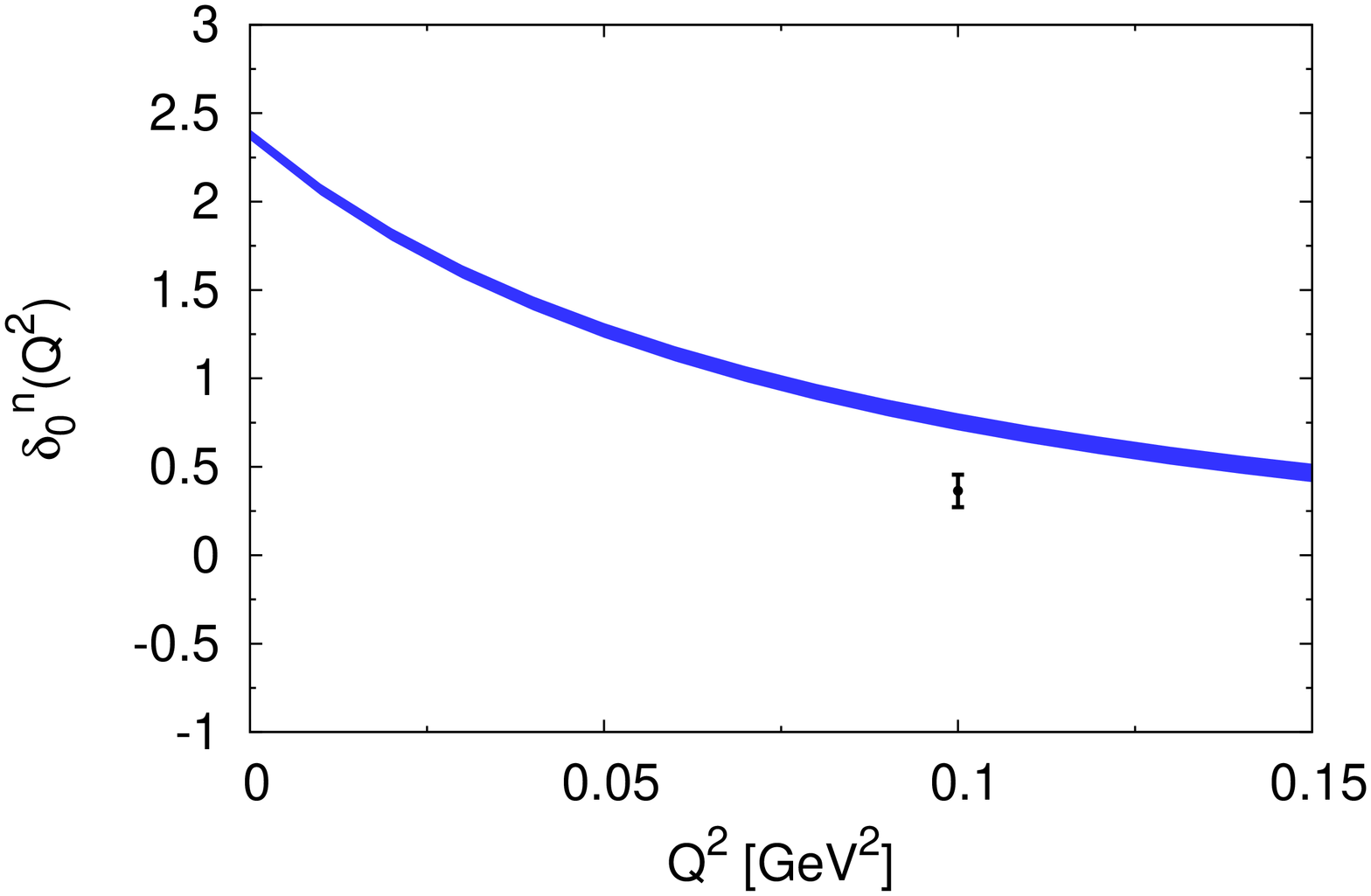}\hspace{0.5cm}
\includegraphics[width=0.33\textwidth,angle=270]{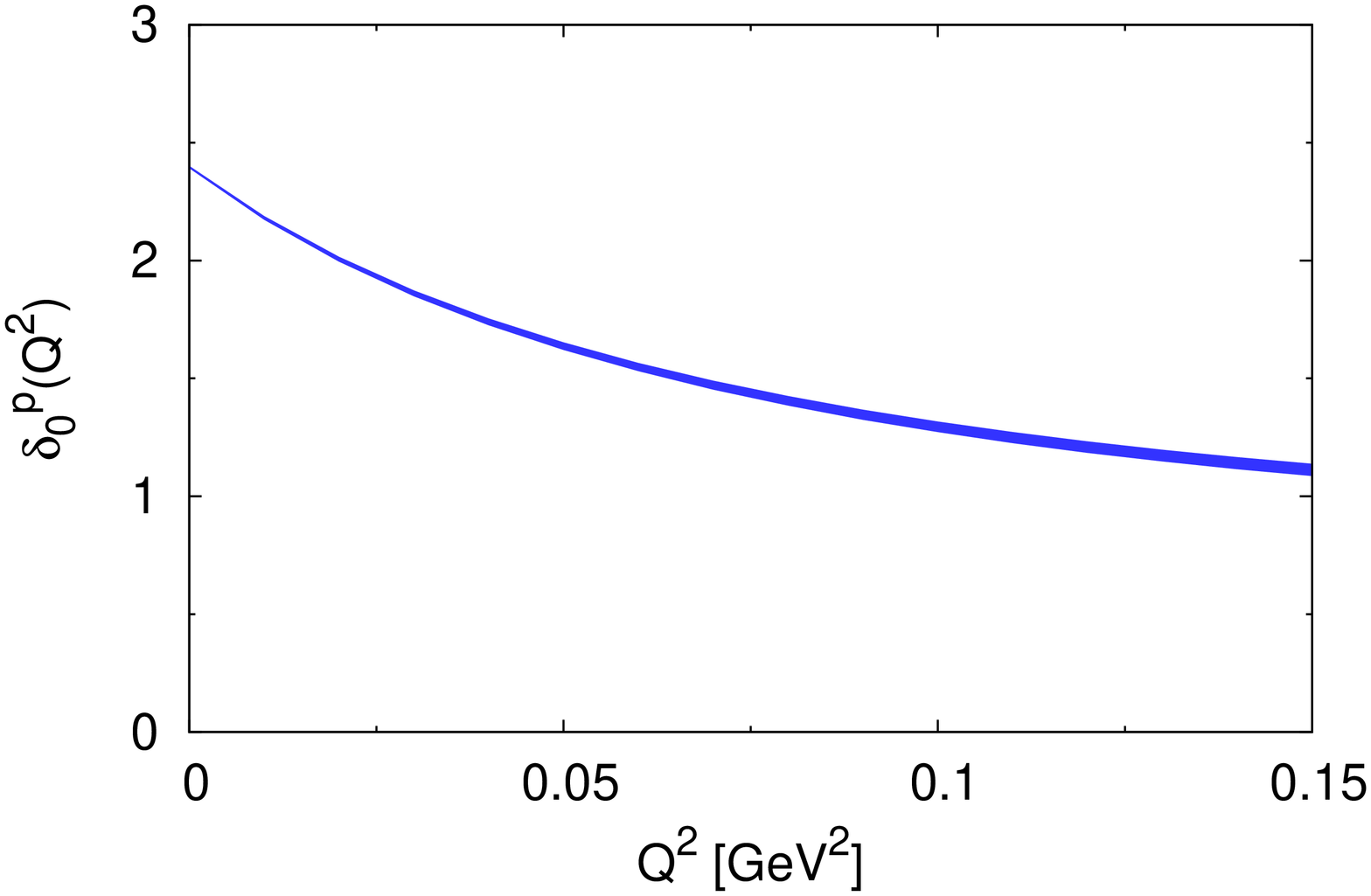}
\end{center}
\caption{Longitudinal-transverse spin polarizability 
in units of $10^{-4}$ fm$^4$ at finite photon  virtuality for the
neutron (left) and the proton (right). Neutron data:
Ref.~\cite{Amarian:2004yf}. Only statistical errors are shown.}

\label{fig:delta0Q2}
\end{figure}

Next, we consider the various observables at finite photon virtuality. In 
Fig.~\ref{fig:gamma0Q2}, we show the neutron (left panel) and proton (right
panel) forward spin polarizabilities for virtualities $Q^2 \leq 0.15\,$GeV$^2$.
For the neutron, there is only one data point at $Q^2 = 0.1\,$GeV$^2$, which
lies slightly above the predictions. The trend of the proton data is not
recovered, the discrepancy between the chiral prediction and the data grows
with increasing photon virtuality. This 
is also reflected in the deviations
of the isoscalar and isovector combinations at $Q^2= 0.1\,$GeV$^2$ given
in Ref.~\cite{Deur:2008ej}.
To get more insights into  these trends, we display the
same decomposition for $\gamma_0^{p,n}(Q^2)$ at $Q^2= 0.1\,$GeV$^2$ as given
at the photon point in Eq.~(\ref{eq:parts}):
\beqa\label{eq:gaq1}
\gamma_0^p&=&1.17_{q^3}\, - \, 4.29_{\epsilon^3, {\rm
    tree}}\, - \, 0.13_{\epsilon^3, {\rm loop}} \, = \, -3.25
~~[\pm 0.48]~,\nonumber \\
\gamma_0^n&=&1.49_{q^3}\, - \, 4.29_{\epsilon^3, {\rm
    tree}}\, - \, 0.15_{\epsilon^3, {\rm loop}} \, = \, -2.95
~~[\pm 0.48]~.
\eeqa
We see that the decrease of  $\gamma_0^{p,n}(Q^2)$ is a combined
effect of a decreasing positive contribution of the pion-nucleon
loops and an increase in magnitude of negative contribution from
the $\Delta$ tree graphs.  Again, a complete fourth calculations is
required to settle the issue. 
Given that the prediction for $\gamma_0^p$ at the
photon point is already close to experiment, one  may hope that such
a fourth order calculation  would provide a  fine test of the chiral
QCD dynamics in view of the upcoming data at low photon virtualities
from Jefferson Lab (down to $Q^2 \approx 0.01\,$GeV$^2$).
It is also interesting to confront our predictions
with the isospin separated forward spin-polarizabilities of
Ref.~\cite{Deur:2008ej}. At $Q^2=0.1\, {\rm GeV}^2$, these authors find
$\gamma_0^{p-n} = 1.53$ and  $\gamma_0^{p+n} = -2.51$. This should
be compared with our predictions of $\gamma_0^{p-n} = -0.30$
and  $\gamma_0^{p+n} = -6.20$ for the central values (all in canonical units). 
The third order SSE calculation disagrees markedly from the experimental values.
As stressed before, a complete ${\cal O}(\varepsilon^4)$ calculation is called for.

The $Q^2$-dependence of the transverse-longitudinal spin polarizability 
is shown in Fig.~\ref{fig:delta0Q2}. In contrast to the IR calculation 
of Ref.~\cite{Bernard:2002pw}, it is a monotonically decreasing function of
$Q^2$ for both the neutron and the proton. This is a generic feature
of using dimensional instead of infrared regularization.
Still, the only empirical value of
$\delta_0^n (Q^2= 0.1\,{\rm GeV}^2)$ is slightly missed  by the chiral
prediction, but again a clear improvement as compared to earlier calculations
is achieved. We remark  again that the uncertainty shown here does not
involve the effects of higher orders not considered here.
 The decrease of $\delta_0^{p,n}(Q^2)$ is mostly due to the
pion-nucleon loop graphs, cf. Eq.~(\ref{eq:parts})  
\beqa\label{eq:deq1}
\delta_0^p&=&0.59_{q^3}\, - \, 0.53_{\epsilon^3, {\rm
    tree}}\, + \, 1.23_{\epsilon^3, {\rm loop}} \, = \,\,\,\,\,\, 1.29
~~[\pm 0.03]~~,\nonumber \\
\delta_0^n&=&0.95_{q^3}\, - \, 0.53_{\epsilon^3, {\rm
    tree}}\, + \, 0.33_{\epsilon^3, {\rm loop}} \, = \,\,\,\,\,\, 0.75
~~[\pm 0.05]~.
\eeqa

\begin{figure}[t!]
\begin{center}
\includegraphics[width=0.33\textwidth,angle=270]{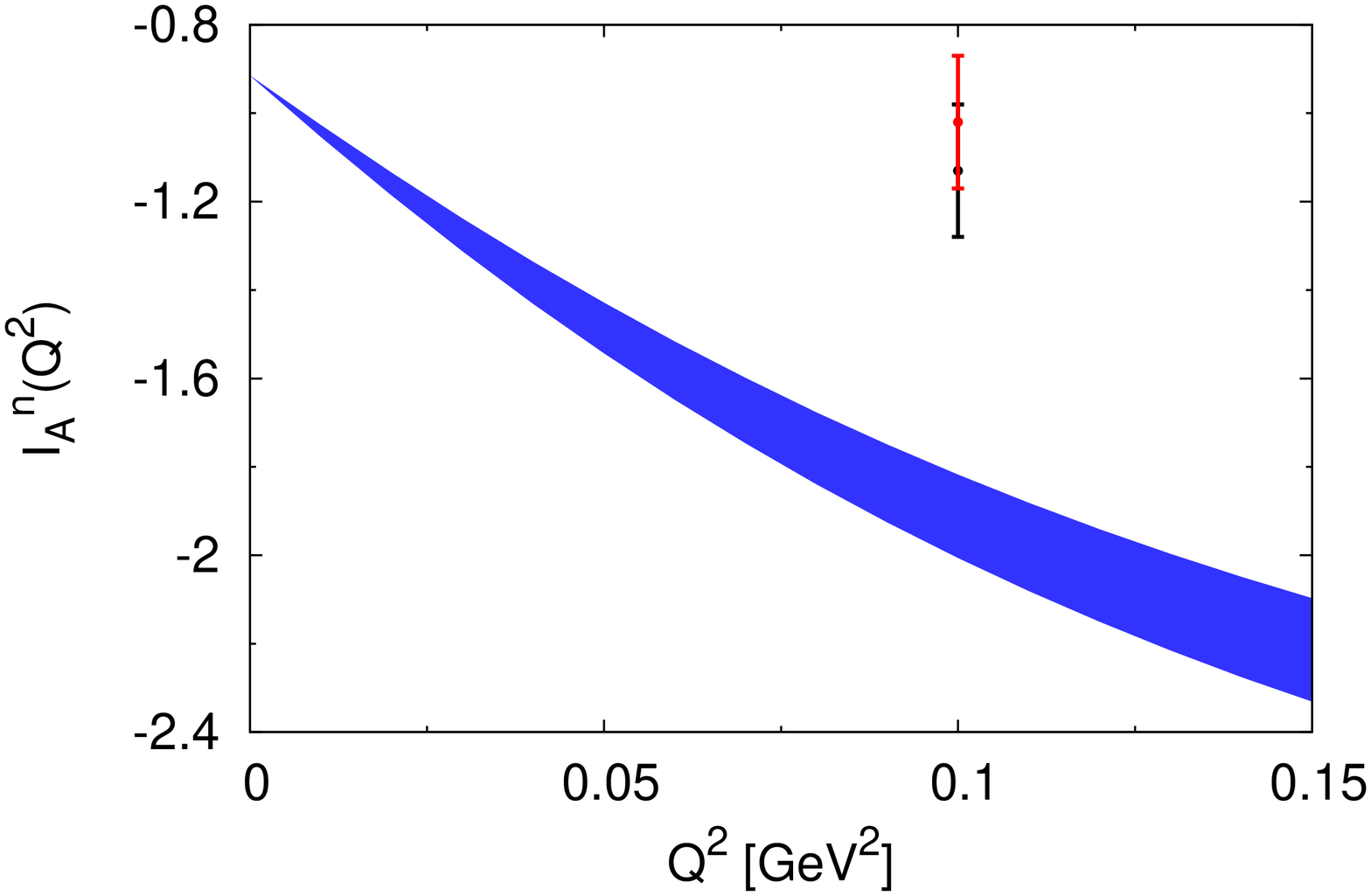}\hspace{0.5cm}
\includegraphics[width=0.33\textwidth,angle=270]{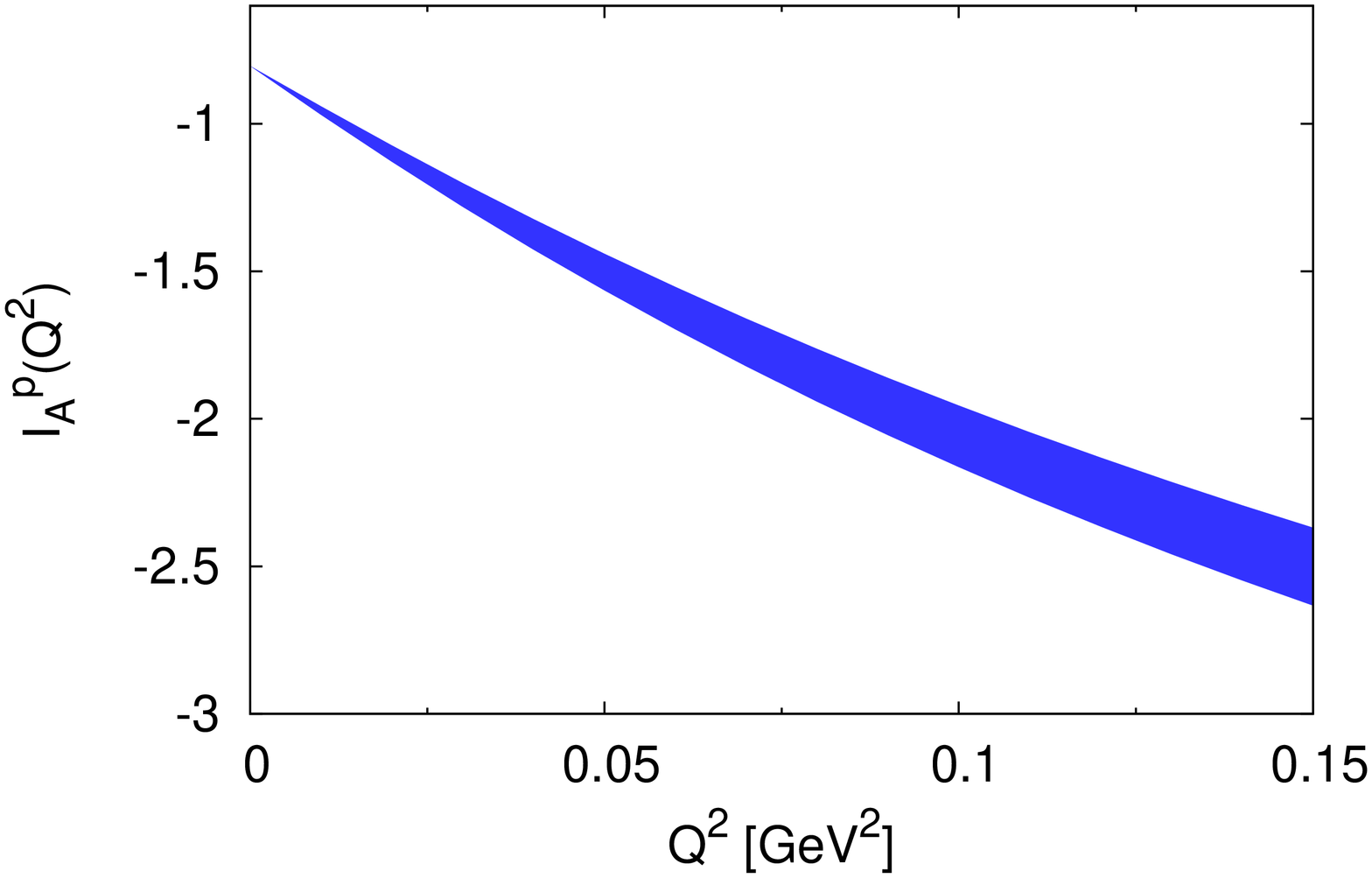}
\end{center}
\caption{Generalized GDH integral $I_A$  for the
neutron (left) and the proton (right). 
Neutron data: Ref.~\cite{Amarian:2002ar}. 
The two data points at the same value of $Q^2$ refer to different extraction methods 
as described in~\cite{Amarian:2002ar}. Only the dominant systematic errors are
shown.}
\label{fig:IAQ2}
\end{figure}

The $Q^2$-dependence of the generalized GDH sum rule $I_A (Q^2)$ is
shown in Fig.~\ref{fig:IAQ2}.  Here, we find a clear difference to the
data point at $Q^2 = 0.1\,$GeV$^2$, whereas the phenomenological
inclusion of the tree level $\Delta$-terms produced a broad band that was
consistent with this datum. It remains to be seen how the complete
one-loop calculation will do, as we know that there are sizeable
${\cal O}(q^4)$ pion-nucleon loop corrections.

\begin{figure}[h!]
\begin{center}
\includegraphics[width=0.33\textwidth,angle=270]{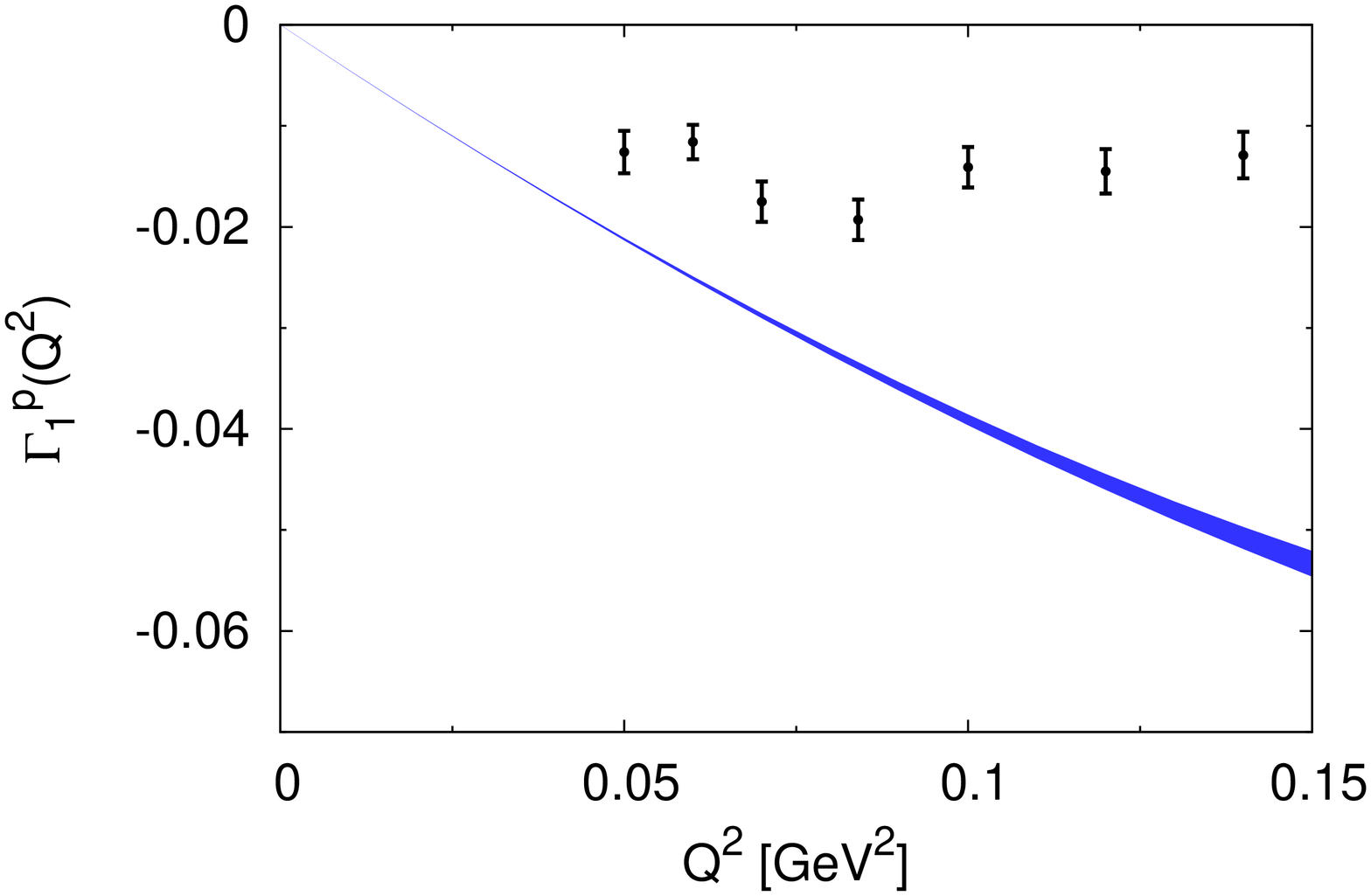}\hspace{0.5cm}
\includegraphics[width=0.33\textwidth,angle=270]{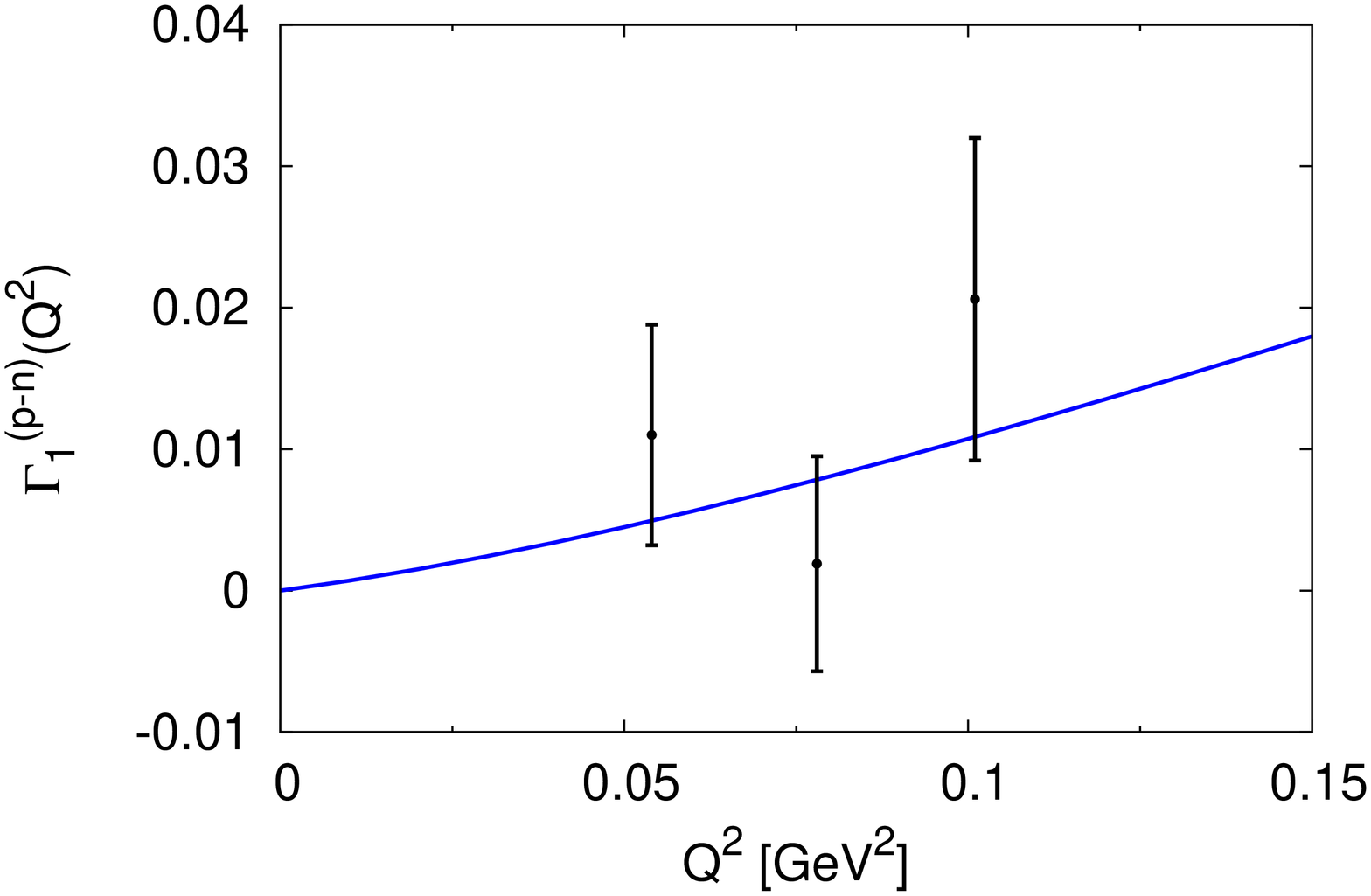}
\end{center}
\caption{First moment of the integral $I_1(Q^2)$ for the
the proton (left) and the isovector nucleon $(p-n)$ (right). 
The proton data are from Ref.~\cite{Prok:2008ev} and the
isovector data from Ref.~\cite{Deur:2008ej}. Only statistical errors are shown.}
\label{fig:Gamma1Q2}
\end{figure}

Finally, the  $Q^2$-dependence of the first moment $\Gamma_1 (Q^2)$
for the proton and the isovector combination $\Gamma_1^{(p-n)}(Q^2)$
are displayed in Fig.~\ref{fig:Gamma1Q2}. While the ${\cal O}(\varepsilon^3)$
contributions slightly improve the chiral prediction for the proton, more curvature
from the pion-nucleon and pion-delta loop graphs at fourth order is
required. This again points towards the necessity of performing such a
complete fourth order calculations within the framework outlined here.
However, we note that the third order calculation already describes
the admittedly relatively imprecise data for the isovector combination
$\Gamma_1^{(p-n)}(Q^2)$ taken from Ref.~\cite{Deur:2008ej}. Therefore,
in this combination the fourth order corrections should largely cancel,
which was found to be the case in the heavy baryon approach \cite{Ji:1999mr}
but not in the infrared regularized covariant calculation \cite{Bernard:2002bs}.

\section{Summary and outlook}
\label{sec:summ}

We have presented a calculation of the nucleon spin structure
at low energies in the framework of a covariant formulation of
baryon chiral effective field theory
with explicit spin-3/2 degrees of freedom. We have included all
terms up-to-and-including ${\cal O}(\varepsilon^3)$ in the small
scale expansion. At this order, one has contributions from
delta tree diagrams supplemented by  the leading pion-nucleon 
and pion-delta loop graphs. Having fixed the LECs from the strong 
and the  electromagnetic width of the $\Delta$-resonance, we can make 
parameter-free predictions. This is particularly
useful in view of the upcoming precision data from Jefferson Lab at low
photon virtualities. The main results
of this investigation can be summarized as follows:
\begin{itemize}
\item[i)] We find an improved description of the forward spin polarizability
$\gamma_0$ for the neutron and the proton. In particular, the value of $\gamma_0^p(0)$ is
consistent with the determination from the GDH collaboration. 
However, the $Q^2$-dependence of the $\gamma_0^p(Q^2)$ is not consistent with the data,
with the discrepancy increasing with larger photon virtuality.
\item[ii)] For the longitudinal-transverse spin polarizability, we find
an improved description as compared to earlier calculations. Still, the
experimental value of $\delta_0^n (Q^2=0.1\,{\rm GeV}^2)$ is slightly larger
in magnitude than the chiral prediction.
\item[iii)] The generalized GDH integral $I_A(Q^2)$ shows a faster fall-off
with increasing photon virtuality as indicated by the data on the neutron.
$\pi N$ and $\pi \Delta$ loops at fourth order are expected to supply the
necessary curvature.
\item[iv)] Similar statements can be made for the first moment $\Gamma_1
(Q^2)$, where the approximate $Q^2$-independence of the proton data 
(for small values of $Q^2$) is not captured by the third order calculation. 
However, we find that the prediction for the isovector combination
 $\Gamma_1^{(p-n)}(Q^2)$ is in agreement with the few existing but not
very precise  data. 
\end{itemize} 
All this points towards the necessity of performing a  complete one-loop
calculation \cite{BEKMprep}. However, we would like to stress that what
was considered here are just the leading contributions based on a 
covariant effective Lagrangian with explicit deltas -- as such, most
of the results can be considered quite encouraging.
In view of the upcoming Jefferson Lab 
data at very small photon virtualities, one can finally hope to test 
the chiral QCD dynamics related to
the nucleon spin structure with sufficient precision. 

\bigskip\bigskip

\section*{Acknowledgments}

\smallskip\noindent

We thank Marina Dorati for collaboration during the
early stage of this investigation and Jian-ping Chen for
strong encouragement to write up these results.
Work supported in part by DFG (SFB/TR 16,
``Subnuclear Structure of Matter''), by the European
Community-Research Infrastructure Integrating Activity ``Study of
Strongly Interacting Matter'' (acronym HadronPhysics3, 
Grant Agreement n. 283286) under the Seventh Framework Programme of EU,
and and ERC project 259218 NUCLEAREFT.

\bigskip

\appendix
\section{Modifying the $\Delta$-resonance input}

So far, we have used the Breit-Wigner mass for the $\Delta$, 
$m_\Delta = 1232\,$MeV and the corresponding width, 
as it was done e.g. in Ref.~\cite{Bernard:2002pw}.
However, one might alternatively use the parameters from the
S-matrix pole deduced from pion-nucleon scattering and pion
photoproduction, which are currently listed as 
\beq
m_\Delta = 1210~{\rm MeV}~, \quad -2\, {\rm Im}~\Sigma_\Delta 
= \Gamma_\Delta^{\rm  str} + \Gamma_\Delta^{\rm  em} = (100 \pm 2)~{\rm MeV}~.
\eeq
Using these values, the couplings $h_A$ and $b_1$ change to
\beq
h_A = 1.51 \pm 0.02~, ~~~~ b_1 = -(5.10\pm 0.27)/m_N~.
\eeq
These are consistent within uncertainties with the earlier
values, cf. Eq.~(\ref{eq:delcoups}), but we note that the
central value for $b_1$ has increased by about 3\%. Using this
new input, we have repeated the calculation. Here, we only
show the modified results for the spin-polarizabilities $\gamma_0$
and $\delta_0$ at the photon point and one finite photon virtuality.
We find at $Q^2=0$:
\beqa
\gamma_0^p&=&2.07_{q^3}\, - \, 4.52_{\epsilon^3, {\rm
    tree}}\, - \, 0.22_{\epsilon^3, {\rm loop}} \, = \, -2.67
~[\pm 0.49]~,\nonumber \\
\gamma_0^n&=&3.06_{q^3}\, - \, 4.52_{\epsilon^3, {\rm
    tree}}\, - \, 0.23_{\epsilon^3, {\rm loop}} \, = \, -1.69
~[\pm 0.48]~,\nonumber\\
\delta_0^p&=&1.54_{q^3}\, - \, 0.42_{\epsilon^3, {\rm
    tree}}\, + \, 1.40_{\epsilon^3, {\rm loop}} \, = \,\,\,\,\,\, 2.52
~[\pm 0.01]~,\nonumber \\
\delta_0^n&=&2.41_{q^3}\, - \, 0.42_{\epsilon^3, {\rm
    tree}}\, + \, 0.37_{\epsilon^3, {\rm loop}} \, = \,\,\,\,\,\, 2.36
~[\pm 0.04]~.
\eeqa
While there are only small changes in the delta-loop contributions, the
delta-tree terms are markedly enlarged, which is in particular relevant
for $\gamma_0$. This increase is approximately to one third due to the increased
value of $b_1$ and to two thirds related to the smaller $\Delta$-mass in
denominator, cf. Eq.~(\ref{eq:born}) (see also Eq.~(39) in Ref.~\cite{Kao:2002cp}).  
Also consistent with the heavy baryon results \cite{Kao:2002cp}, the
corrections to $\delta_0$ are less significant. These trends persist  
at finite photon virtualities. At $Q^2=0.1$~GeV$^2$, we find
\beqa
\gamma_0^p&=&1.17_{q^3}\, - \, 5.52_{\epsilon^3, {\rm
    tree}}\, - \, 0.17_{\epsilon^3, {\rm loop}} \, = \, -4.52
~~[\pm 0.60]~,\nonumber \\
\gamma_0^n&=&1.49_{q^3}\, - \, 5.52_{\epsilon^3, {\rm
    tree}}\, - \, 0.20_{\epsilon^3, {\rm loop}} \, = \, -4.23
~~[\pm 0.60]~,\nonumber\\
\delta_0^p&=&0.59_{q^3}\, - \, 0.65_{\epsilon^3, {\rm
    tree}}\, + \, 1.42_{\epsilon^3, {\rm loop}} \, = \,\,\,\,\,\, 1.36
~~[\pm 0.03]~,\nonumber \\
\delta_0^n&=&0.95_{q^3}\, - \, 0.65_{\epsilon^3, {\rm
    tree}}\, + \, 0.37_{\epsilon^3, {\rm loop}} \, = \,\,\,\,\,\, 0.67
~~[\pm 0.06]~.
\eeqa
These data show the same trends as at the photon point, cf. Eqs.~(\ref{eq:gaq1},\ref{eq:deq1}).
As we are only considering the leading delta tree and loop graphs here, we expect
that some of the uncertainty induced by the values for $m_\Delta, \Gamma_\Delta$ will be
reduced when the subleading ${\cal O}(\varepsilon^4)$ corrections are included.



\vskip 1cm
\begin{thebibliography}{99}

\frenchspacing

\bibitem{Kuhn:2008sy}
  S.~E.~Kuhn, J.-P.~Chen and E.~Leader,
  Prog.\ Part.\ Nucl.\ Phys.\  {\bf 63} (2009) 1
  [arXiv:0812.3535 [hep-ph]].


\bibitem{Amarian:2002ar} 
  M.~Amarian, L.~Auerbach, T.~Averett, J.~Berthot, P.~Bertin, W.~Bertozzi, T.~Black and E.~Brash {\it et al.},
  Phys.\ Rev.\ Lett.\  {\bf 89}, 242301 (2002)
  [nucl-ex/0205020].

\bibitem{Amarian:2004yf}
  M.~Amarian {\it et al.}  [Jefferson Lab E94010 Collaboration],
  Phys.\ Rev.\ Lett.\  {\bf 93} (2004) 152301
  [nucl-ex/0406005].

\bibitem{Prok:2008ev}
  Y.~Prok {\it et al.}  [CLAS Collaboration],
  Phys.\ Lett.\  B {\bf 672} (2009) 12
  [arXiv:0802.2232 [nucl-ex]].

\bibitem{Chen:2010qc}
  J.~P.~Chen,
  Int.\ J.\ Mod.\ Phys.\ E {\bf 19} (2010) 1893
  [arXiv:1001.3898 [nucl-ex]].

\bibitem{CD12}
V.~Sulkosky,
  ``An Overview of Longitudinal Spin Structure Measurements from JLab,''
  talk at Chiral Dynamics 2012, Jefferson Lab., August 2012.

\bibitem{Bernard:2007zu}
  V.~Bernard,
  Prog.\ Part.\ Nucl.\ Phys.\  {\bf 60}, 82 (2008)
  [arXiv:0706.0312 [hep-ph]].

\bibitem{Kao:2002cp}
  C.~W.~Kao, T.~Spitzenberg and M.~Vanderhaeghen,
  Phys.\ Rev.\  D {\bf 67}, 016001 (2003)
  [arXiv:hep-ph/0209241].

\bibitem{Bernard:2002bs}
  V.~Bernard, T.~R.~Hemmert and U.-G.~Mei\ss ner,
  Phys.\ Lett.\  B {\bf 545}, 105 (2002)
  [arXiv:hep-ph/0203167].

\bibitem{Bernard:2002pw}
  V.~Bernard, T.~R.~Hemmert and U.-G.~Mei\ss ner,
  Phys.\ Rev.\  D {\bf 67}, 076008 (2003)
  [arXiv:hep-ph/0212033].

\bibitem{Edelmann:1998bz}
  J.~Edelmann, N.~Kaiser, G.~Piller and W.~Weise,
  Nucl.\ Phys.\ A {\bf 641} (1998) 119
  [nucl-th/9806096].


\bibitem{Ji:1999mr}
  X.~D.~Ji and J.~Osborne,
  J.\ Phys.\ G {\bf 27}, 127 (2001)
  [arXiv:hep-ph/9905410].

\bibitem{Hemmert:1997ye}
  T.~R.~Hemmert, B.~R.~Holstein and J.~Kambor,
  J.\ Phys.\ G {\bf 24}, 1831 (1998)
  [arXiv:hep-ph/9712496].

\bibitem{Bernard:2005fy}
  V.~Bernard, T.~R.~Hemmert and U.-G.~Mei\ss ner,
  Phys.\ Lett.\  B {\bf 622} (2005) 141
  [arXiv:hep-lat/0503022].

\bibitem{Vermaseren:2000nd}
  J.~A.~M.~Vermaseren,
  arXiv:math-ph/0010025.

\bibitem{Passarino:1978jh}
  G.~Passarino and M.~J.~G.~Veltman,
  Nucl.\ Phys.\ B {\bf 160} (1979) 151.


\bibitem{Dutz:2003mm}
  H.~Dutz {\it et al.}  [GDH Collaboration],
  Phys.\ Rev.\ Lett.\  {\bf 91} (2003) 192001.

\bibitem{Deur:2008ej}
  A.~Deur, P.~Bosted, V.~Burkert, D.~Crabb, V.~Dharmawardane, G.~E.~Dodge, T.~A.~Forest and K.~A.~Griffioen {\it et al.},
  Phys.\ Rev.\ D {\bf 78} (2008) 032001
  [arXiv:0802.3198 [nucl-ex]].

\bibitem{BEKMprep}
  V.~Bernard, E.~Epelbaum, H.~Krebs and U.-G.~Mei\ss ner,
  {\it in preparation}.




\end{thebibliography}
\end{document}